\documentstyle[psfig]{mn}
\title{Pulsar motions in our Galaxy}
\author[X.~H. Sun \& J.~L. Han]
{X.~H. Sun and J.~L. Han \\
National Astronomical Observatories, Chinese Academy of Sciences, 
Beijing 100012, China\\
xhsun@bao.ac.cn~(XHS), hjl@bao.ac.cn~(JLH)
        }

\date{Accepted \hspace{2cm};
      Received \hspace{2cm} 
     }
\pubyear{2003}

\begin{document}

\maketitle
\label{firstpage}

\begin{abstract}
Pulsar motions in our Galaxy from their birth until 2 Gyr are studied
statistically via Monte-Carlo simulation of $2\times 10^5$ pulsars
with the best available representation of the Galactic potential.  We
find that the distribution of height above the Galactic plane
for pulsars with characteristic ages less than about 8~Myr could be well
fitted by a Gaussian function. For older pulsars, an extra exponential
function is necessary to fit the distribution. The scale-height of the
Gaussian component increases linearly with time until about 40~Myr.
The height distribution becomes stabilized after about 200~Myr.  These
results are not sensitive to initial height or radial
distributions. Taking the relationship between the initial velocity and
height distribution, we found from the latest pulsar catalog that the
height distribution of pulsars younger than 1~Myr directly implies the
mean initial velocity of $280\pm96$~km~s$^{-1}$. Comparison of
simulated sample of pulsars with the current available millisecond
pulsars shows that their 1D initial velocity dispersion should be 
most probably $60\pm10$~km~s$^{-1}$.
\end{abstract}


\section{Introduction}
Pulsars are high velocity objects in our Galaxy \cite{ll94}.  They were born
in supernova explosions near the Galactic plane where their progenitors
live, but move away very fast from the plane.  The evolution of pulsar
heights was considered \cite{bwh+92,hbw+97,mk97} but not explicitly
described in most pulsar population syntheses, therefore, the picture for
pulsar height evolution was not very clear. We want to demonstrate in this
paper by 3D simulations how pulsars move in our Galaxy.

Birth velocities of both normal and millisecond pulsars (MSPs) have
been investigated by many authors. For normal pulsars, Lyne \&
Lorimer~\shortcite{ll94} studied transverse speeds of 29 pulsars younger
than 3~Myr and obtained the mean birth velocity $\bar{V_{\rm
B}}=$450$\pm$90~km~s$^{-1}$. Taking into account selection effects and
using all available pulsar proper motion data in population synthesis, Lorimer,
Bailes \& Harrison~\shortcite{lbh97}  found the mean birth velocity
$\bar{V_{\rm B}} \sim$500~km~s$^{-1}$. Hansen \&
Phinney~\shortcite{hp97} considered the distribution of proper
motions, including all available upper limits, and
concluded that the mean birth velocity $\bar{V_{\rm B}}$ should be smaller,
only 250-300~km~s$^{-1}$. Cordes \& Chernoff~\shortcite{cc98} performed a
detailed analysis for measured velocities of 49 young pulsars.
They favoured a two-component Gaussian model in 3-dimension with
characteristic velocities\footnote{In the Maxwellian distribution, velocity
in each dimension is a Gaussian distribution.  The mean velocity in 3-D and
velocity dispersion in 1-D are connected by $\bar{V_{\rm B}} \sim \sigma_v
\sqrt{8/\pi}\simeq 1.6 \sigma_v$.} $\sigma_v$ of 175~km~s$^{-1}$ and
700~km~s$^{-1}$.  Arzoumanian, Chernoff \& Cordes~\shortcite{acc02} infer
the velocity distribution of radio pulsars detected in 400-MHz surveys, taking
into account beaming, selection effects and luminosity evolution. They
also favour a two-component Gaussian model with $\sigma_v \sim
90$~km~s$^{-1}$ and 500~km~s$^{-1}$.  

For MSPs, Lorimer~\shortcite{lor95}
accounted for known survey selection effects and obtained the local surface
density of MSPs, birthrate and the lower limit of the mean birth velocity
that is $\sigma_v > 50$km~s$^{-1}$, i.e. $\bar{V_B}\geq80$km~s$^{-1}$.  The
results were confirmed by Cordes \& Chernoff~\shortcite{cc97}, who obtained
the velocity perpendicular to the Galactic plane,
$v_z=52^{+17}_{-11}$km~s$^{-1}$, i.e.~a 3-D velocity dispersion $\sigma_v=
84$~km~s$^{-1}$ using likelihood analysis on previous surveys for MSPs plus
selection effects. Later, Lyne et al.~\shortcite{lml+98} found that the
population syntheses of the MSPs with the Maxwellian initial 1D velocity
dispersion of $80\pm20$~km s$^{-1}$ agrees the best with the data of newly
discovered MSPs in Parkes southern sky survey together with previous MSPs.
The 3D mean birth velocity is correspondingly $130\pm30$~km~s$^{-1}$.  These
results were later confirmed by Toscano et al.~\shortcite{tsb+99} using
more MSP proper motions obtained from timing measurements.

Previously, Narayan \& Ostriker~\shortcite{no90} tried to model the observed
pulsar populations. They deduced an analytic formula for the height
distribution by assuming a Gaussian distribution at all times and solving the
energy conservation equation in $z$-direction (perpendicular to the Galactic
plane). However, the height distribution changes from time to time when
pulsars run away from the Galactic plane. The integrated $z$-distribution of
pulsars with different ages may not be a Gaussian. We will show by our 
Monte-Carlo simulation that the scale-height evolution is much more complicated
than a Gaussian.  Analytic solutions obtained from the over-simplified form
of a Galactic acceleration model can hardly represent the realistic situation,
especially over long time with non-linear evolution of pulsar heights. By
simulations, we are able to track the height distribution of a pulsar sample,
for any form of acceleration.

In our work, the initial velocity dispersion for MSPs is obtained by
comparison of the height distribution between simulated and observed pulsar
samples. We also tried to derive the initial velocity dispersion of normal
pulsars by comparing their scale-heights and characteristic ages. In Section 2
we will describe the simulation procedures together with input parameters
and governing equations. Several possibilities of parameter choice are also
discussed. In Section 3, we show the simulation results, mostly regarding
how the pulsar scale-height evolves. We demonstrate in Section 4 two
applications of our simulations, i.e.~to determine the birth
velocity dispersions of normal pulsars and MSPs. The influence of the
selection effects is also discussed. The conclusions are given in Section 5.

\section{Details of simulation}

In our simulations, we take a Galactocentric rectangular coordinate system,
where the x and y-axes are orthogonal in the Galactic plane and z-axis
perpendicular to the plane. The dynamic status of a pulsar at any time can
be described by $(x,y,z;v_x,v_y,v_z)$, where $x$, $y$ and $z$ are position
coordinates, and $v_x$, $v_y$ and $v_z$ the velocities along each axis.
Obviously, the Galactocentric radius $R=\sqrt{x^2+y^2}$. The distance between
the Sun and the Galactic center is taken as $R_0=8.0$ kpc.
Simulation results should not depend on coordinate system taken. Before
calculating pulsar positions at any time $t$, the initial positions and
velocities of pulsars and the Galactic acceleration should be discussed.

\subsection{Initial height distribution}

Initial positions of pulsars should follow the positions of their
progenitors, e.g.~OB stars, or the distribution of supernova
remnants. Ma\'iz-Apell\'aniz \shortcite{mai01} has listed all previous
determinations of scale heights of the OB star disk in either an exponential
or a Gaussian height distribution. He obtained a solid measurement of
the vertical distribution of B-B5 stars in solar neighborhood as
\begin{equation} 
P_z(z_{\rm ini}) =\frac{1}{\sqrt{2\pi}h_{\rm ini}}
\exp(-\frac{z_{\rm ini}^2}{2h_{\rm ini}^2})
\end{equation}
with a scale-height of $h_{\rm ini}=63$~pc from {\it Hipparcos} data, though
a single-component, self-gravitating, and isothermal disk model with a
$sech$ function is as good as the Gaussian.  Here we ignore the halo
component which is about 4\% of stars, and it is not clear whether this
component is related to MSPs. In our simulations, we assume this height
distribution to be adequate for any places in the disk.

We also tried an exponential distribution $ P_z(z_{\rm ini}) = (1/2h_{\rm
ini}) \exp(-|z_{\rm ini}|/h_{\rm ini}) $ with a scale-height $h_{\rm
ini}=60$~pc and a flat distribution in infinitely thin disk ($z_{\rm
ini}=0$). We found that {\it our results are not sensitive to different
initial $z$-distributions}, as all of them {\it lead to the same or very
similar dynamic behavior} after one million year.  In the following we will
use the Gaussian distribution in Eq.(1), and assume it is valid at all
Galactocentric radii.

\subsection{Initial radial distribution}

\begin{figure}  
\psfig{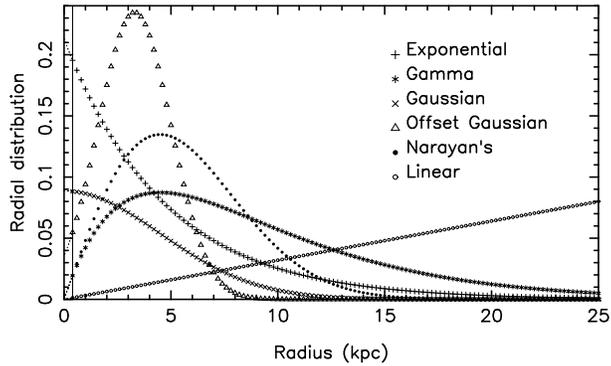}
\caption[]{Different forms of radial distribution 
  plotted against radius $R$. A uniform density distribution
  is a {\it linearly} increasing function $P_R$ against $R$.
  The vertical line at $R=0.4$ kpc indicates the minimum $R$ in
  our simulation.
}
\label{Rini_distr}
\end{figure}

The radial density distribution of newly born neutron stars is not clear at
all. Narayan~\shortcite{nar87} made a Gaussian fit to the observed radial
density profile obtained by Lyne, Manchester \& Taylor~\shortcite{lmt85} as
using a normalized function
\begin{equation}
\rho_R(R)=\frac{1}{64\pi}\exp[-(\frac{R}{8})^2] {\rm (kpc)^{-2}},
\label{nr}
\end{equation}
where $\rho_R$ is radial density. Bailes \& Kniffen~\shortcite{bk92} pointed
out that another function with $\rho_R\to 0$ as $R\to 0$ is also consistent
with the observed data. However, most of the subsequent work took Narayan's
fit as a starting distribution in $R$ (e.g. Lorimer et al. 1993; Johnston
1994; Mukherjee \& Kembhavi 1997\nocite{lbd+93,joh94,mk97}). While, this
distribution reflects the observed distribution in $R$, rather than the
initial distribution which most probably is related to the exponential
distribution of OB stars in the Galactic plane, as expected in our Galaxy
and other spiral galaxies~\cite{bah86,pac90}.  Note here that {\it the radial
probability distribution} ($P_R$), which is used for discussions below and 
plots in Figure~\ref{Rini_distr}, should have the form of
\begin{equation}
P_R \; dR=\frac{2\pi R  \; \rho_R\ dR}{\int_0^{\infty}2\pi R \;  \rho_R \ dR}.
\end{equation}
For Narayan's function\footnote{We note that in literature some authors
claimed to use the Gaussian radial distribution of
Narayan~\shortcite{nar87}, but as a mistake, they used the form of
$\rho_R(R)$ to represent $P_R$. The prefactor of $P_R$, $R/32$, is $R$
dependent, while that of $\rho_R(R)$ is a constant of $1/(64\pi)$.},
$P_R=\frac{R}{32}\exp[-(\frac{R}{8})^2]$. In fact, this function has a 
natural deficit for $R \to 0$ which can be evidently seen from the observed
data that Johnston~\shortcite{joh94} tried to explain.

In our simulations, we have tested a few probability functions, as
shown in Figure.~\ref{Rini_distr}.\\
{\bf (1)} {\it The exponential distribution}: 
\begin{equation}
P_R(R_{\rm ini})=\frac{1}{R_{\rm exp}} 
\exp(-\frac{R_{\rm ini}}{R_{\rm exp}}),
\end{equation}
hereafter $R_{\rm ini}=\sqrt{x_{\rm ini}^2+y_{\rm ini}^2}$ is the initial
Galactocentric radius. The re-scaled characteristic radius
$R_{\exp}=4.7$~kpc\footnote{The authors implicitly used $R_0=8.5$ kpc in
context. Here we scaled it to $R_0=8.0$~kpc.  Same are the following cases 3,
4.}, as given in Hartman et al.(1997\nocite{hbw+97});\\
{\bf (2)} {\it The Gamma distribution}:
\begin{equation}
P_R(R_{\rm ini})=a_R\; \frac{R_{\rm ini}}{R_{\exp}^2} 
\exp(-\frac{R_{\rm ini}}{R_{\rm exp}}),
\label{rmax}
\end{equation}
with $R_{\exp}=4.5$~kpc and 0.4~kpc$\leq R_{\rm ini}\leq 25$~kpc following
Paczy\'{n}ski~\shortcite{pac90}.  This radial distribution corresponds to
exponential radial density. Here $a_R=1.0683$. See also Gonthier et
al.~\shortcite{gob+02};\\
{\bf (3)} {\it The Gaussian distribution},
\begin{equation}
P_R(R_{\rm ini})= \frac{1}{\sqrt{2\pi}R_{\rm g}}
\exp(-\frac{R_{\rm ini}^2}{2R_{\rm g}^2}),
\end{equation}
with a rescaled $R_{\rm g}=4.5$~kpc following Lorimer et 
al.~\shortcite{lbd+93}. 
Many authors (e.g. Hartman et al. 1997\nocite{hbw+97}) used this distribution 
afterwards;\\
{\bf (4)} {\it The offset Gaussian distribution}: 
\begin{equation}
P_R(R_{\rm ini})=\frac{1}{\sqrt{2\pi}R_{\rm g}}
 \exp[-\frac{(R_{\rm ini}-R_{\rm off})^2}{2R_{\rm g}^2}],
\end{equation}
with rescaled $R_{\rm g}=1.7$~kpc and $R_{\rm off}=3.3$~kpc given 
by Hartman et al.~\shortcite{hbw+97};\\
{\bf (5)} {\it The distribution from Narayan}:
\begin{equation}
P_R(R_{\rm ini})=\frac{R_{\rm ini}}{R_{\rm g}^2}
\exp(-\frac{R_{\rm ini}^2}{2R_{\rm g}^2})
\end{equation}
here $R_{\rm g}=4.5$~kpc taken from the normalized factor of the Gaussian
distribution;\\
{\bf (6)} {\it The linear distribution}:\\
\begin{equation}
P_R(R_{\rm ini})=\frac{2R_{\rm ini}}{R_2^2-R_1^2}
\end{equation}
so that in a given range of Galactocentric radii
$R_1<R_{\rm ini}<R_2$ (we took $R_1=0.4$~kpc and $R_2=25$~kpc) 
the pulsar surface density to be a constant.\\
We have simulated pulsar motions in all above cases and will make
comparison on the evolved radial distribution, while for detailed
studies, we will concentrate on the Gamma distribution (Eq.~\ref{rmax}).

\subsection{Initial velocities}

The initial velocities can be written as $\vec{v}_{\rm ini}=\vec{v}_{\rm
birth}+\vec{v}_{\rm rot}$, where $\vec{v}_{\rm birth}$ is the birth velocity
and $\vec{v}_{\rm rot}$ the rotation velocity of the Galaxy. The physical
origin of the birth velocity is not clear hitherto, which might be generated
from the disruption of a binary~\cite{go70}, rocket effects~\cite{ht77} or
asymmetric supernova explosion~\cite{dc87}. We will not go further into details
but assume that the birth velocity is isotropic and follows a Maxwellian
distribution
\begin{equation}
P_{v}(v_{\rm birth})=\sqrt{\frac{2}{\pi}} 
\frac{v_{\rm birth}^2}{\sigma_{\rm birth}^3}
\exp(-\frac{v_{\rm birth}^2}{2\sigma_{\rm birth}^2}),
\end{equation}
where $\sigma_{\rm birth}$ is 1D velocity dispersion, corresponding to the
mean velocity of $\sqrt{8/\pi}\sigma_{\rm birth}$ or a 3D velocity
dispersion of $\sqrt{3}\sigma_{\rm birth}$. We will show the dynamical
results of $\sigma_{\rm birth}=$100, 200, 300 and 400 km~s$^{-1}$ in our
simulations\footnote{Note here 1~km~s$^{-1}\approx1$~pc~Myr$^{-1}$.} .

\subsection{Galactic potential and acceleration}
Pulsars accelerate in the Galactic gravitational potential, as the
acceleration being $\vec{g}=-\nabla\phi.$ The potential $\phi$ is determined
by the mass distribution in our Galaxy, $\nabla^2\phi=4\pi G\rho, $ where
$G$ is the gravitational constant, and $\rho$ mass density as a function
of position. One can find that
\begin{equation} 
\nabla\cdot\vec{g}=-4\pi G\rho 
\label{grho}
\end{equation}
In fact, Eq.~\ref{grho} can be rewritten as 
\begin{equation}
\rho=-\frac{1}{4\pi G}\left\{\frac{\partial g_z}{\partial z}
+\frac{1}{R}\frac{\partial\left(R\;g_R\right)}{\partial R}\right\}
\label{rhogg}
\end{equation} 
in a cylindrical coordinate system. Here $g_z$ is the acceleration in
$z$ direction and $g_R$ in $R$ direction. 
In order to simulate pulsar dynamics, it is very important to find
the best representation of the Galactic potential.

\subsubsection{Adopted Galactic potential model}
We used the potential model of disk/bulge originally proposed by Miyamoto \&
Nagai \shortcite{mn75},
\begin{equation}
\phi_i\left(R,z\right)=\frac{GM_i}{
\left\{R^2+\left[a_i+\left(z^2+b_i^2\right)
^{1/2}\right]^2\right\}^{1/2}},
\end{equation} 
here $a_i$, $b_i$ and $M_i$ are model parameters, and $i=1$ stands for the
bulge and $i=2$ for the disk. Using the density distribution of the halo
given by Kuijken \& Gilmore~\shortcite{kg89},
\begin{equation}
\rho=\frac{\rho_c}{1+(r/r_c)^2}.
\end{equation} 
here $r^2=x^2+y^2+z^2$, Paczy\'nski~\shortcite{pac90} derived the potential
for the halo component as
\begin{equation}
\phi_h=-\frac{GM_c}{r_c}\left[\frac{1}{2}\ln\left(1+\frac{r^2}{r_c^2}\right)
+\frac{r_c}{r}{\rm atan}\frac{r}{r_c}\right],
\end{equation}
where $M_c=4\pi\rho_cr_c^3$ and $\rho_c$ and $r_c$ are model parameters.
Paczy\'nski~\shortcite{pac90} determined all the parameters (see Table
\ref{par_pac}) for the disk, bulge and halo potentials to make a good agreement
with the rotation curve, local volume density and the column density between
$z=\pm700$~pc.  He took $R_0=8.0$ kpc. This potential expression has been
used for simulations of pulsar motions by Hansen \& Phinney (1997), Cordes
\& Chernoff (1998) and Gonthier et al. (2002) \nocite{hp97,cc98,gob+02}.
Since the rotation curve for $R<0.4$ kpc can not be well described by
the formula of the potentials, as Sofue \& Rubin (2001) show, all our
simulations have been limited to $R>0.4$ kpc.

\begin{table}  
\centering
\caption[]{Potential parameters given by Paczy\'nski~
\shortcite{pac90} with subscripts 1 and 2 corresponding to 
bulge and disk and c to halo.}\label{par_pac}
\begin{tabular}{cc}
\hline\hline
a$_1$~(kpc)        &    0 \\
b$_1$~(kpc)        &    0.277\\
M$_1$~(M$_\odot$)  &    1.12$\times10^{10}$\\
a$_2$~(kpc)        &    3.7 \\
b$_2$~(kpc)        &    0.20\\
M$_2$~(M$_\odot$)  &    8.07$\times10^{10}$\\
r$_c$~(kpc)        &    6.0\\
M$_c$~(M$_\odot$)  &    5.0$\times10^{10}$\\\hline\hline
\end{tabular}
\end{table}

\subsubsection{Galactic potential previously used}
Previously, several models for Galactic mass distribution and 1D
acceleration have been used to simulate pulsar motions in the $z$-direction. 

{\bf 1.} Narayan \& Ostriker~\shortcite{no90} modeled the mass density in
the Galaxy as a thin layer at $z=0$ with surface density $\Sigma$ and a
uniform halo with density of $\rho$, and obtained the acceleration in
$z$-direction as,
\begin{equation} 
g_z=4\pi G\rho\left[z+\frac{\Sigma}{2\rho}\frac{|z|}{z}\right].
\end{equation}
This model was also used by Itoh \& Hiraki \shortcite{ih94} and tried by
Hartman \& Verbunt \shortcite{hv95}.

\begin{figure}  
\psfig{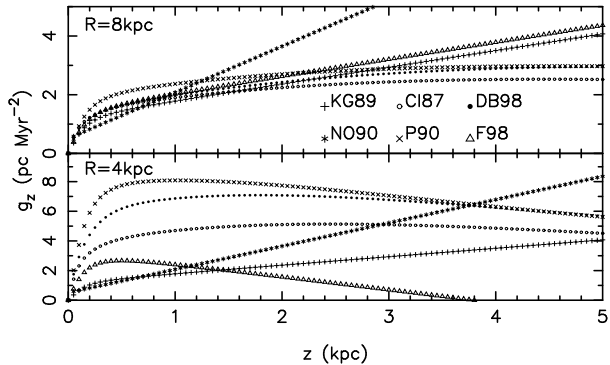}
\caption[]{Acceleration $g_z$ calculated for $R=4$~kpc and 8 kpc for
the models by Carlberg \& Innanen (1987: CI87), Dehnen \& Binney (1998: DB98),
Ferri\`{e}re (1998: F98), Kuijken \& Gilmore (1989: KG89), Narayan \&
Ostriker (1990: NO90) and Paczy\'{n}ski (1990: P90).  }
\label{gz_z}
\end{figure}

{\bf 2.} Kuijken \& Gilmore~\shortcite{kg89} rewrote
Eq.(\ref{rhogg}) as
\begin{equation}
\rho=-\frac{1}{4\pi G}\left\{\frac{\partial g_z}{\partial z} +
2(A^2-B^2)\right\}
\end{equation} 
for disk galaxies following Mihalas \& Binney (1981). Here $A$ and $B$ are
the Oort constants. After considering the double-exponential disks and the
spherical components (halo, bulge and corona), they modeled the distribution
of matter in the disk via tracer stars at high $z$ and obtained the
acceleration in $z$-direction
\begin{equation}
g_z=1.04\times10^{-3}\left(\frac{1.26z}{\sqrt{z^2+0.18^2}}
+0.58z\right)
\label{kg89}
\end{equation}
which can be further used to determine the disk surface density. 

Their formula of $g_z$ has been used by Bhattacharya et al.~(1992)
and Hartman \& Verbunt (1995) to simulate the pulsar motions
in $z$-directions for investigation of pulsar magnetic field
decay, and extended by Ferri\`ere (1998) for ISM distribution
studies. It seems to have been much better constrained at
high $z$ than that of Narayan \& Ostriker~\shortcite{no90}.
See Figure~\ref{gz_z} and discussions by Hartman \& Verbunt (1995).

\begin{figure}
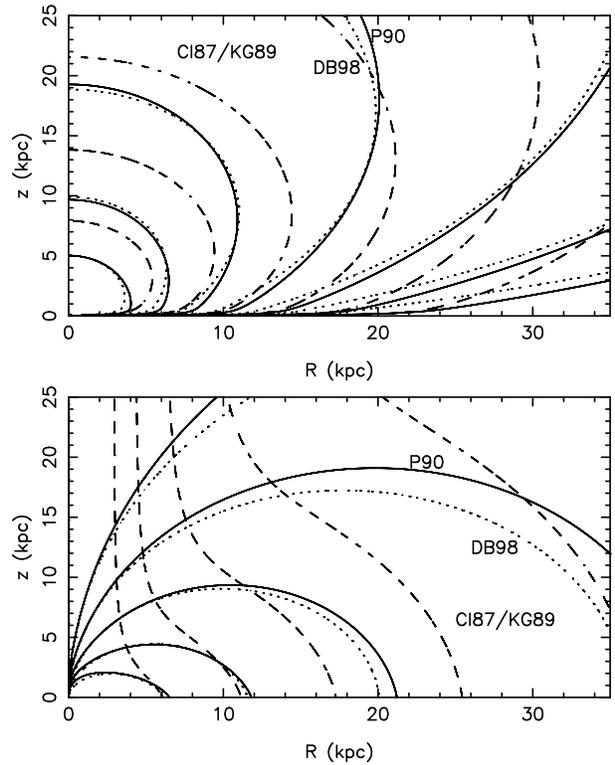
  
\psfig{file=Fig3a.ps,width=8cm,angle=-90}
\psfig{file=Fig3b.ps,width=8cm,angle=-90}
\caption[]{The contours of $g_z$ (upper panel) and $g_R$ (lower panel)
calculated from potentials of Dehnen \& Binney (1998: DB98), Paczy\'{n}ski
(1990: P90), and Carlberg \& Innanen (1987) with modified parameters by
Kuijken \& Gilmore (1989: CI87/KG89).  The contour levels are at 2$^n$ pc
Myr$^{-2}$, with n= $0,\; \pm1,\; \pm2,\; \pm3$, and with notes near n=0.}
\label{gcont}
\end{figure}

{\bf 3.}  The potential model for the disk/spheroid and
nucleus/bulge, either directly from Carlberg \& Innanen (1987) or using
parameters given by Kuijken \& Gilmore~\shortcite{kg89}, has the form of
\begin{equation}
\phi_{dh}=\frac{GM}{\left\{\left[a+\sum_{i=1}^3\beta_i\left(z^2+h_i^2
\right)^{1/2}\right]^2+b^2+R^2\right\}^{1/2}},
\end{equation}
\begin{equation}
\phi_{b,n}=\frac{GM_{b,n}}{(b_{b,n}^2+R^2)^{1/2}},
\label{ph}
\end{equation}
which has been used by Lorimer et al.~(1993, 1997), Hartman et al. (1997)
and Mukherjee \& Kembhavi (1997) for pulsar population synthesis.  The model
parameters $a$, $b$, $h_i$, $\beta_i$ and mass $M$ for different components
were constrained by the rotation curve data. 

{\bf 4.}  Dehnen \& Binney (1998\nocite{db98}) have constructed a set of
mass models for our Galaxy using all available observational constraints.
All of their models consist of the spheroidal bulge, halo and three disk
components, namely the thin and thick stellar disks and the interstellar
medium disk.  The density for bulge and halo in the model is described by
the spheroidal density distribution
\begin{equation} 
\rho_s=\rho_0 \left(\frac{m}{r_0}\right)^{-\gamma}
\left(1+\frac{m}{r_0}\right)^{\gamma-\beta}  \exp(-m^2/r_t^2).
\end{equation} 
Here $m=(R^2+q^{-2}z^2)^{1/2}$,
and the density of disks is given by 
\begin{equation} 
\rho_d(R,z) = \frac{\Sigma_d}{2z_d} 
\exp(-\frac{R_m}{R_d} - \frac{R}{R_d} - \frac{|z|}{z_d}),
\end{equation} 
where
$\rho_0$,~$r_0$,~$q$,~$\gamma$,~$\beta$,~$r_t$,~$\Sigma_d$,~$z_d$,~$R_d$ and
$R_m$ are model parameters. One set of the best fitted parameters which we
tried in this paper was derived by satisfying the total local surface density
of 52.1~M$_\odot$~pc$^{-2}$ at the solar neighborhood and circular velocity
$v_c(R_0)=217$~km~s$^{-1}$ with $R_0=8$~kpc (Model 2 in their Table 3). 

\subsubsection{Discussions on Galactic potential models}

Obviously most models of $g_z$ (see Figure~\ref{gz_z}) are acceptable in the
regions {\it only around the sun and near the mid-plane}, which can {\it
not} be used to simulate pulsar motions farther than a few kpc from the Sun
in our Galaxy.  The change of acceleration (both $g_z$ and $g_R$) with $R$
and $z$ (see Figure~\ref{gcont}) must be considered. Previous results based
on simulations with only $g_z$ acceleration (e.g. Bhattacharya et al. 
1992~\nocite{bwh+92}) therefore should be treated with caution. 
\begin{figure}  
\psfig{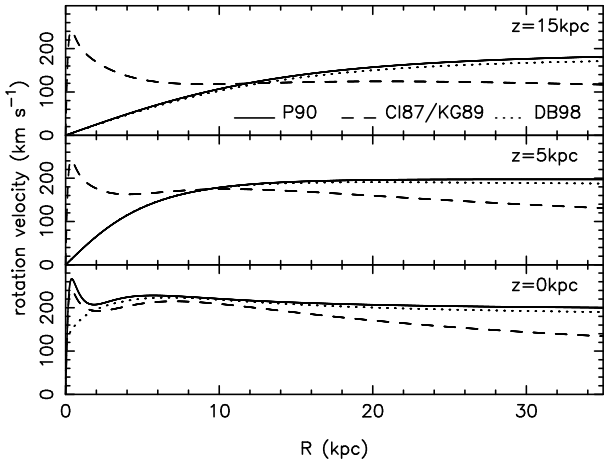}
\caption[]{The rotation curves at different $z$ from the mass
models of Dehnen \& Binney (1998: DB98), Paczy\'{n}ski (1990: P90), 
and Carlberg \& Innanen (1987) with modified parameters by Kuijken
\& Gilmore (1989: CI87/KG89).}
\label{rotv}
\end{figure}

An accurate model for the 
mass distribution and acceleration is needed for better
understanding of pulsar motions in our Galaxy. The rotation curves for
models of Dehnen \& Binney (1998), Paczy\'{n}ski (1990), and Carlberg \&
Innanen (1987) with modified parameters by Kuijken \& Gilmore (1989) are
shown in Figure~\ref{rotv}. We noticed that the potential given by Dehnen \&
Binney (1998) is almost the same with that by Paczy\'{n}ski (1990), though
newer constraints were used.  Our simulations for both potentials
produced very similar results. The main difference between these two models
is the rotation curve peak at $R\sim300$~pc, which Dehnen \& Binney (1998)
did not but Paczy\'{n}ski (1990) did consider the rotation curve given by
Burton \& Gordon (1976) in small-$R$ part fitted from the HI data of
Westerbout (1976). Although the peak at the small-$R$ is not unusual in
spiral galaxies (Sofue et al. 1999), it is not clear whether and how the
central bar affects the rotation in our Galaxy (Dehnen private
communication). Therefore, in our simulations we only consider pulsar
motions in the region of 0.4~kpc$<R<$25~kpc.

We also noticed that the mass model of Carlberg \& Innanen (1987) even with
modified parameters by Kuijken \& Gilmore (1989) cannot produce the proper
rotation curve at high $z$ and large $R$. The very slow change of the
potential near the Galactic center~(Figure~\ref{gcont}) is not reasonable.
The accelerations given by Dehnen \& Binney (1998) and Paczy\'{n}ski (1990)
at small $R$ and at high $z$ (see Figure~\ref{gcont}) seems to be better 
constrained by observational data (see Figure~\ref{rotv}). Therefore the
models of  Dehnen \& Binney (1998) and Paczy\'{n}ski (1990) are recommended 
here. Paczy\'nski's potential is preferable due to its simple analytic form.
 
\subsection{Governing equations and calculation method}

The initial velocities and positions of the pulsars were generated randomly
according to the discussions above. The rotation velocity was calculated
from the Galactic potential. The motion of a pulsar is therefore only
governed by the Newtonian kinetic equation as,
\begin{equation}
\left\{\begin{array}{rcl} \displaystyle{\frac{{\rm
d}\vec{r}\left(t\right)}{{\rm d}t}}&\,=\,&\vec{v} \left(\vec{r}\right)
\\[2mm] \displaystyle{\frac{{\rm d}\vec{v}\left(\vec{r}\right)}{{\rm
d}t}}&\,=\,& \vec{g}\left(\vec{r}\right),
\end{array}\right.
\end{equation}
where $\vec{r}\left(t\right)$, $\vec{v}\left(\vec{r}\right)$ and
$\vec{g}\left(\vec{r}\right)$ are 3D vectors.  The equations above were
solved numerically in three directions using the {\it 5}th order Runge-Kutta
method with adaptive control stepsizes [using the subroutine {\sl rkqs}]
\cite{ptv+92}. The initial stepsize is $dt=10^{-4}$ Myr, and calculations of
position and velocity in 3D then continues until 0.1 Myr with adjusted
stepsizes according to the required accuracy. The data are recorded every
0.1~Myr and the total energy $E_{\rm tot}=(v_x^2+v_y^2+v_z^2)/2+\phi(R,z)$
are checked. The position and velocity then are used for input of the next
0.1 Myr. The fluctuation of the total energy of each pulsar is always
$<\sim2\%$ from 0 to 2~Gyr, which we believe are good enough for simulation
accuracy.

\begin{figure*}
\psfig{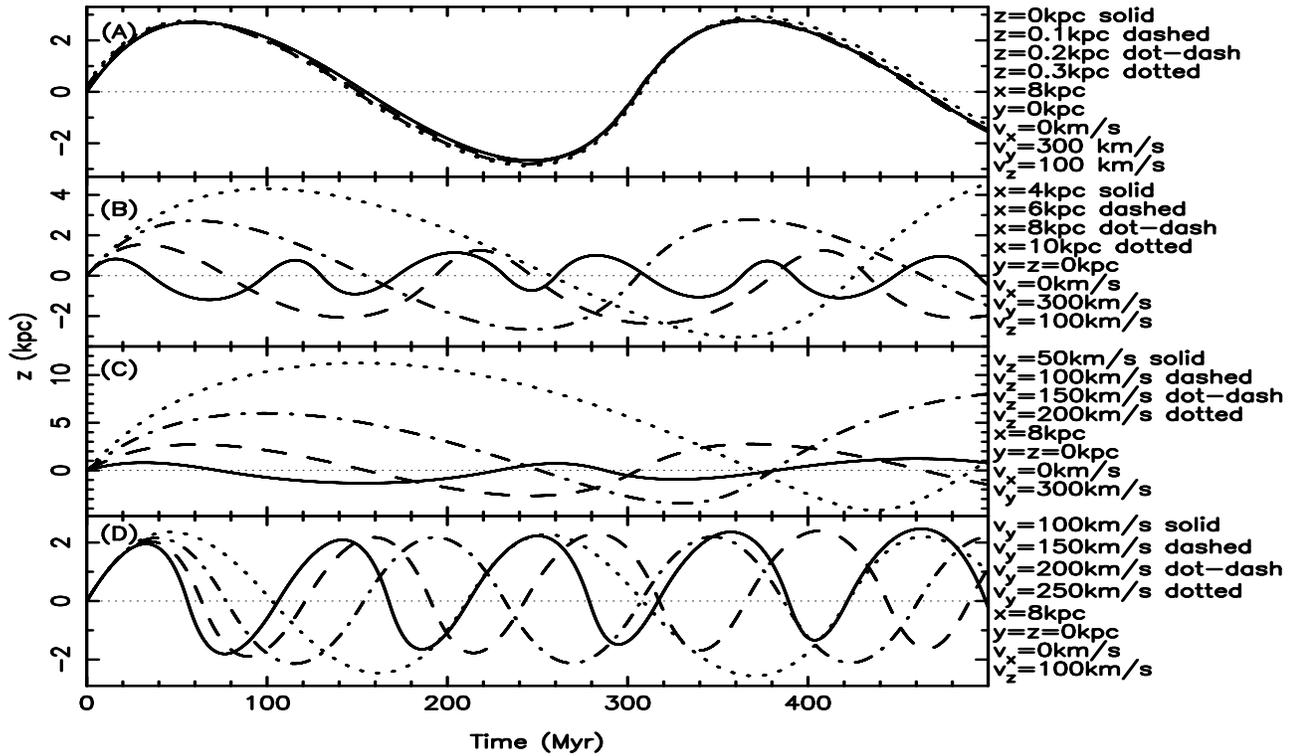}
\caption{The $z$-tracks of pulsars, corresponding 
initial values are marked on the right side of the plots.}
\label{z_track}
\end{figure*}

Some examples of moving trajectories of pulsars are plotted in
Figure~\ref{z_track}, with various initial conditions. Obviously, the
different initial heights have very little influence on $z$ tracks
(Figure~\ref{z_track}-A), while if the pulsar is located at different $R$,
trajectories can be very different (Figure~\ref{z_track}-B). Certainly,
different rotation velocities (Figure~\ref{z_track}-C) or vertical
velocities (Figure~\ref{z_track}-D) will naturally cause different
trajectories.

\section{Results and Discussion}
Limited by the computation resources available, we simulated $2\times10^5$
pulsars at the age $t=0$, and put their initial positions and velocities
according to the details discussed in Section 2. We then traced their motions
up to an age of 2 Gyr, and perform statistics on their locations every 0.1 Myr.
We first present results from Gaussian distributions of the initial height
$z$, and Gamma-distributions for initial $R$, which will be called the
standard initial conditions hereafter. We then compared those with the
results from other initial conditions.

\subsection{Results from standard initial conditions}

\subsubsection{The scale-height evolution}
The $z$-heights of simulated pulsars are binned with equal height interval
every 0.1~Myr. We used two data-sets: the average height of the {\it i}th
bin, $z(i)$, and the number of pulsars in the {\it i}th bin, $N(i)$.  The
Levenburg-Marquardt method was then employed to fit these two data-sets with
given functions (p.678 of Press et al. 1992)\nocite{ptv+92} to obtain the
height distribution.

\begin{figure}
\psfig{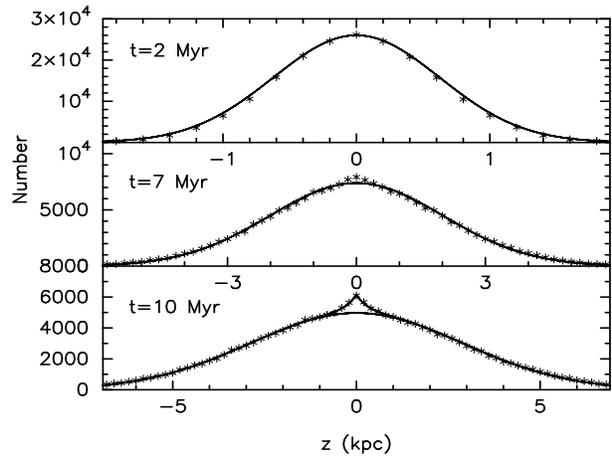}
\caption[]{The height distribution can be well fitted by a Gaussian
(t$<$8~Myr) or a Gaussian plus an exponential functions (t$>$8~Myr). Here
$\sigma_{\rm birth}=300$~km s$^{-1}$ is assumed.}
\label{heightfit}
\end{figure}

\begin{figure}
\psfig{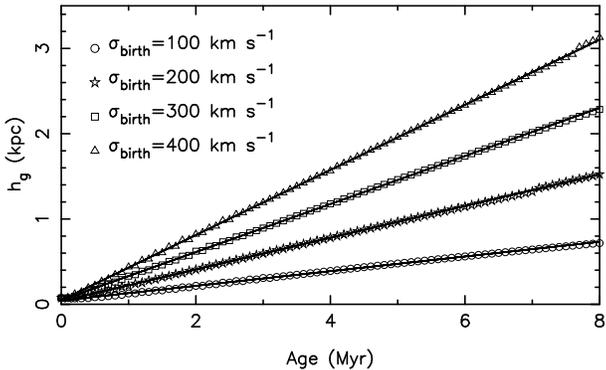}
\caption[]{The scale-heights of the Gaussian components ($h_g$) increases
linearly with ages, in spite of different birth velocities.}
\label{hg_fitting} 
\end{figure}

For pulsars with an age less than 8~Myr, the height distribution can be well
fitted by a single Gaussian function (Figure~\ref{heightfit}: t=2~Myr and
7~Myr), i.e.,
\begin{equation}
N(i)=A\exp(-\frac{z^2(i)}{2h_g^2}).
\end{equation}
Here $h_g$ is the scale-height and $A$ is amplitude. We found that $h_g$
increases linearly with $t$~(Figure~\ref{hg_fitting}), which can be
represented as
\begin{equation}
h_g=h_0+\sigma t,
\label{hg_t}
\end{equation} 
where $h_0$ and $\sigma$ are fitting coefficients listed in
Table~\ref{par_fit}.


\begin{table}
\centering
\caption[]{Fitting parameters for different initial 1D velocity dispersion}
\label{par_fit}
\begin{tabular}{ccc}
\hline
$\sigma_{\rm birth}$   &  $h_0$ &   $\sigma$ \\
   km s$^{-1}$   &   pc   &  km s$^{-1}$\\
\hline
     100           &   45   &    86 \\
     200           &   37   &   186 \\
     300           &   49   &   282 \\
     400           &   48   &   382 \\
\hline
\end{tabular}
\end{table}

\begin{figure}
\psfig{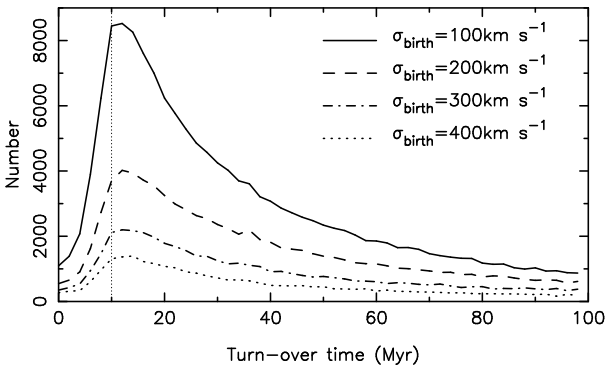}
\caption[]{The distribution of pulsar turn-over times.}
\label{return_time}
\end{figure}

The relation between $\sigma$ and $\sigma_{\rm birth}$ should be discussed.
At small $t$, one could take
\begin{equation}
z_t=z_{\rm ini}+v_{\rm zini} t 
\label{approx}
\end{equation}
as a simplification, where $z_t$ is height at time $t$. To justify this
approximation, we calculated the turn-over time of pulsars, which was
defined as the time when pulsars change the sign of $v_z$, i.e.~start to
move towards the Galactic plane for the first time. It is approximately a
quarter of the oscillation period for a pulsar moving up and down. As can be
seen from Figure~\ref{return_time}, most pulsars change their motion
directions (i.e.~the sign of $v_z$) at $\sim$15~Myr, a time not sensitive
to initial velocities. Therefore, for a Maxwellian initial 3D velocity
distribution, it is natural that the distribution of height at small age $t$
can have a Gaussian form, and that the scale-height increases with time as
$h_g\sim \sigma_{\rm birth} \ t$. We noticed that roughly $\sigma_{\rm
birth}\sim\sigma$, a relation that can be used to determine 1D initial
velocity dispersions. We emphasized that formula (\ref{approx}) is only an
approximation of the first order. The effect of the deceleration of the
Galactic potential did work {\it to a certain extent}, which made $\sigma$
slightly smaller, i.e.~$\sigma<\sim\sigma_{\rm birth}$~(Table~\ref{par_fit}).

\begin{figure} 
\psfig{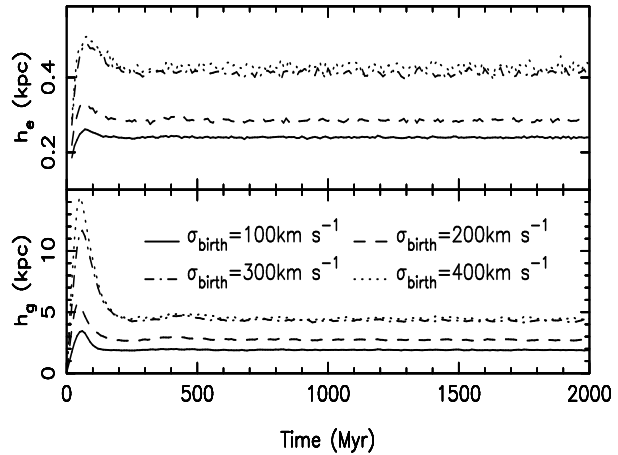}
\caption[]{The evolution of the scale-heights of Gaussian component
($h_g$: lower panel) and exponential component ($h_e$: upper panel).}
\label{hghe_0_2Gyr} 
\end{figure}

As time increases, the central peak in the height distribution gets more and
more prominent (Figure~\ref{heightfit}: t=10~Myr). A single Gaussian
function cannot fit the distribution satisfactorally. 
We found that a Gaussian plus an exponential function provides
a much better fit, i.e.:
\begin{equation}
N(i)=A\exp(-\frac{z^2(i)}{2h_g^2})+B\exp(-\frac{\left|z(i)\right|}{h_e}),
\end{equation}
where $h_g$ and $h_e$ are scale-heights, and $A$ and $B$ are amplitudes.
The exponential component mainly models the smaller
heights and the Gaussian component accounts for the larger ones. 
The scale-height of
Gaussian component, $h_g$, increases linearly with time until $t\sim40$~Myr
(Figure~\ref{hghe_0_2Gyr}). 

When $t>\sim40$~Myr, the height distribution gets more concentrated towards
lower heights and $h_g$ decreases gradually (Figure~\ref{hghe_0_2Gyr}).
When $t>\sim200$~Myr, the height distribution tends to stabilize, and
both $h_g$ and $h_e$ show no large variations. We noticed that the larger
the initial velocity dispersion, the longer the stabilization time
and the larger the resulting scale-height.

\subsubsection{Scale-heights at different radii}

We have shown earlier that the scale-height of $z$ distribution of all
pulsars, from 0.4~kpc$\leq R \leq$25~kpc. It is not clear whether the 
$z$ distribution changes with $R$.

\begin{figure} 
\psfig{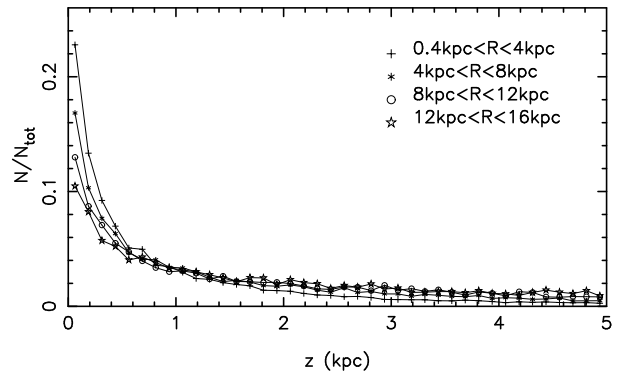}
\caption[]{The height distribution at $t=$ 2~Gyr in 4 $R$-ranges.  The
simulation used $\sigma_{\rm birth}=$  300~km~s$^{-1}$. }
\label{z_distr_R}
\end{figure}

The height distributions in four ranges of $R$ does not show significant
discrepancies~(Figure~\ref{z_distr_R}).

\subsubsection{Evolution of the radial distribution}

\begin{figure}
\psfig{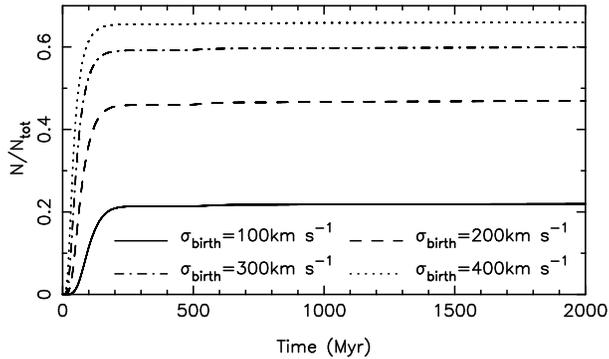}
\caption[]{Fraction of the total number of pulsars that escape the Galaxy.}
\label{Num}
\end{figure}

We assume that all simulated pulsars are initially located at 0.4~kpc$<R<$
25~kpc with the maximum at $R=$4.5~kpc. During the simulation, we trace only
those pulsars in the region 0.4~kpc$\leq R\leq$25~kpc. As mentioned above,
the Galactic potential near the Galactic centre may not be realistic. This
might cause exotic pulsar motion trajectories \cite{ci87} so that we ignore
all pulsars moving close to the centre ($R<$0.4kpc). Pulsars moving out to
$R>25$ kpc are defined as ``escaped'' pulsars, since the Galactic
gravitation is too weak to bound them. The number of escaped pulsars
increases until 200 Myr (Figure~\ref{Num}). The larger the initial
velocities, the more likely the pulsars are to escape. For $\sigma_{\rm birth}=
400$~km~s$^{-1}$, more than 60\% of pulsars escape after 100 Myr.
Obviously a huge amount of pulsars, even after their radio emission
turns off, have gone into Galactic halo or intergalactic space, which
could be a very important ingredient of the dark-matter halo.

\begin{figure} 
\psfig{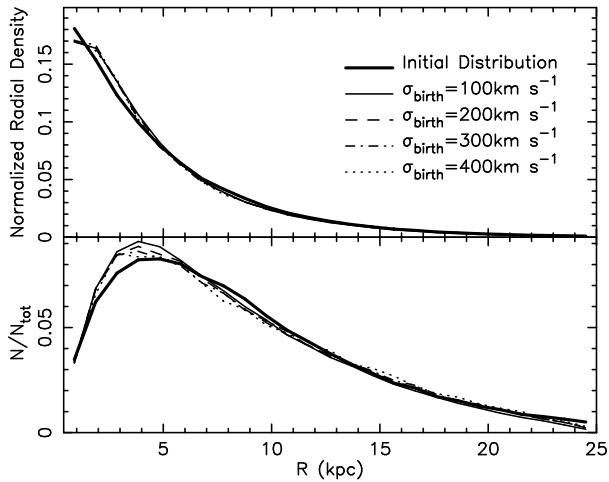}
\caption[]{The $R$-distribution (lower panel) and the normalized radial
density (upper panel) at t=0 (initial) and t=2~Gyr.}
\label{R_distr}
\end{figure}

We found that there is almost {\it no evolution} of the radial density and
radial probability distribution (Figure~\ref{R_distr}) if the initial
radial distribution is a Gamma-distribution. The different velocities do
not change the distribution although numbers of remained pulsars change
a lot.

Concerning the form of
the radial distribution, there is a deficit towards the Galactic centre,
as Johnston~\shortcite{joh94} discussed. A peak appears at $\sim4.5$~kpc, 
not too different from the initial distribution. However, the surface 
density does not have such a deficit, in contrast to the observed pulsar
density distribution newly determined by Lorimer~\shortcite{lor04}.

\subsection{Results for other initial conditions}

\subsubsection{Alternative initial height distributions}
Two kinds of initial height distributions have been tried for simulations of
pulsar motions: an exponential height distribution, and a flat distribution
of all $z_{\rm ini}=0$. 

\begin{figure} 
\psfig{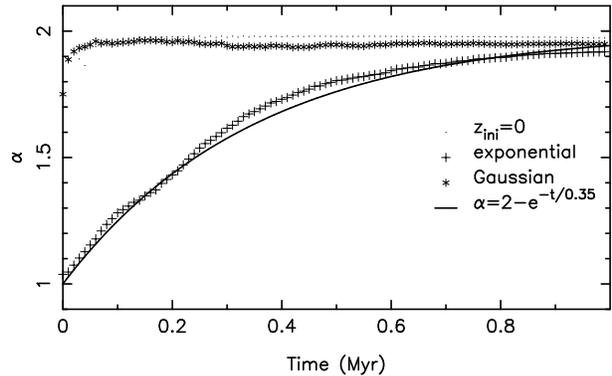}
\caption[]{The change of the index $\alpha$ in a generalized fitting function 
$N(i)=A\exp\{-(\frac{|z(i)|}{h_\alpha})^\alpha\}$.}
\label{alpha}
\end{figure}

For those non-Gaussian initial height distribution, a Gaussian function 
did not always give a good fit when $t<1$~Myr.  So we tried another
function for the height distribution:
\begin{equation}
N(i)=A\exp\left\{-\left(\frac{\left|z(i)\right|}{h_\alpha}\right)^\alpha
\right\}, 
\end{equation}
where $\alpha=1$ corresponds to an exponential function and $\alpha=2$ a
Gaussian function. We examined a few cases of initial height distributions.
As shown in Figure~\ref{alpha}, for the initial Gaussian distribution of
heights or the flat distribution with $z_{\rm ini}=0$, $\alpha$ is always
nearly 2, so that we can always fix $\alpha=2$ in the fit. For the
exponential distribution of initial heights, the index $\alpha$ increases
from 1 to 2 gradually, which can be approximately described by a function
\begin{equation}
\alpha=2-\exp\left\{-\frac{t}{0.35~{\rm Myr}}\right\}.
\end{equation}

When $t$ is greater than 1~Myr, the height distributions can be described
by a Gaussian or a Gaussian plus an exponential function,
as shown in Section 3.1.1. We therefore conclude that the pulsar height
distribution is insensitive to the initial distribution after $t>1$ Myr. 

\subsubsection{Alternative initial radial distributions}

As we introduced in Section 2.1, alternative radial distributions can be
used for simulations: (1) Gamma, (2) Gaussian, (3) offset Gaussian,
(4) exponential, (5) Narayan and (6) uniform density distribution. 

\begin{figure}
\psfig{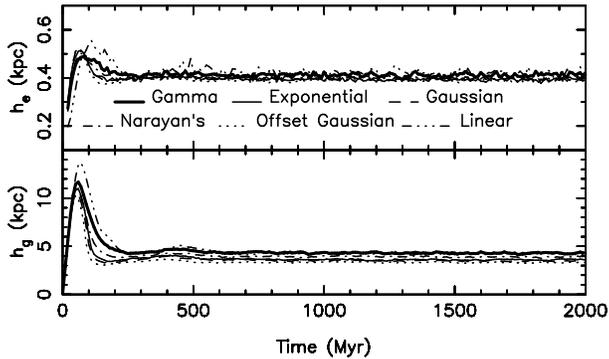}
\caption[]{Evolution of scale-heights from simulations with different radial
distributions.}
\label{hghe_Rdistr}
\end{figure}

We first checked the evolution of height distribution, and found that all
these radial distributions produce similar results, as
Figure~\ref{hghe_Rdistr} shows. That is to say, the height distribution
is insensitive to initial radial distribution.

\begin{figure}
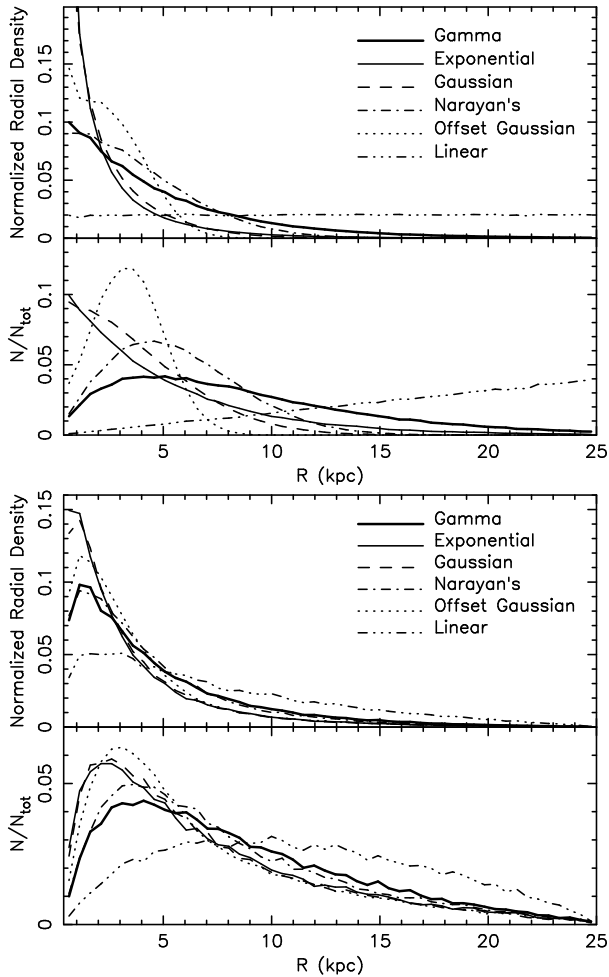

\psfig{file=Fig15a.ps,width=8cm,angle=-90}
\psfig{file=Fig15b.ps,width=8cm,angle=-90}
\caption{The density distribution (normalized $\rho_R$) and the radial
distribution ($P_R$) at t=0 yr (upper panels) and t=2 Gyr (lower
panels) for various initial distributions.}
\label{ev_rd}
\end{figure}

The evolution of radial distributions has been checked using both the density
and radial distributions (see Figure~\ref{ev_rd}). Obviously pulsars at large
$R$ feel a smaller Galactic potential and hence are easy to escape
after some years. The final distribution seems to be more prominent at 
small $R$ where the Galactic potential is much larger due to the presence
of the Galactic bulge.

We indeed see the fall-off at small $R$ in both the density distribution
and radial distribution from our simulations with all kinds of initial
radial distributions. However, compared to the observed deficit at small
$R$ recently determined by Lorimer (2004\nocite{lor04}), the simulated
deficit is much smaller in all cases of radial distributions. Note that
what we simulate simply moving neutron stars and take no account of
their radio emission. It is likely that the beaming effect and the evolution of
pulsar radio emission probably have to be considered carefully for a fair
comparison. We can conclude from our simulation, however, that there should 
be a large number of evolved neutron stars near the Galactic center, which 
may no longer be observable as radio pulsars.

\section{Applications}
All above simulations are made in ideal conditions for pulsar motions
in the Galactic potential. Previous authors have simulated currently
observed pulsar populations, so our results give a complementary image
of pulsars moving in our Galaxy.  As we just mentioned, for detailed
comparison between the simulation results and observed sample, one
has to consider beaming and the evolution of radio emission 
of pulsars. We understand that the selection effects on pulsar surveys
are quite severe for a few reasons. For example, in any flux-limited
survey, more luminous pulsars are easy to be detected, even they are
far away. Faint pulsars can be detected only if they are very nearby.
It is not clear how the luminosity of pulsars evolves with age,
though evidence available shows that young pulsars tend to be brighter.
Due to dispersion smearing, only nearby millisecond pulsars are easy 
to be discovered, mostly at high latitudes.

Nevertheless, at these two age extremes, i.e.~young pulsars and millisecond
pulsars, we do not have to synthesize populations for comparison of the
velocity distribution or height distribution.  As we do not expect the
pulsar luminosity to evolve much in 1 Myr, and also previous surveys have
been mostly concentrated on the Galactic plane where young pulsars live, 
the sample of very young pulsars suffer much less of selections effects in
the surveys. For millisecond pulsars, the height distribution is very stable
after 200~Myr (see Figures~\ref{hghe_0_2Gyr} and \ref{hghe_Rdistr}), so that
millisecond pulsars of any age should follow the same height distribution
fitted by a Gaussian plus an exponential function. In the following we
will discuss the height distribution of young pulsars and millisecond
pulsars.

\subsection{Young pulsars: Initial Velocity Dispersion derived from $z$ 
distribution}

Most pulsar surveys have been conducted near the Galactic plane
where young pulsars are predominantly found. As our simulations show, the
scale-height is linearly increasing for young pulsars in the form of
$h_g=h_0+\sigma t$ for $t<8$ Myr, which may be used to determine the
initial velocity of normal pulsars. 

\begin{figure}  
\psfig{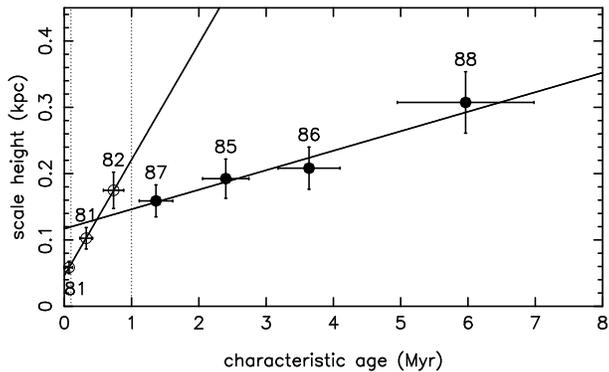}
\caption[]{The scale-heights versus characteristic ages of currently known
pulsars. Solid lines represent linear fits. The numbers of pulsars in 
the corresponding bins are indicated in the plot.}
\label{hg_Norm_NewD}
\end{figure}

We first take all pulsars with characteristic ages of
1~Myr$\le\tau\le$8~Myr from the most updated pulsar
catalog~\cite{mhth04}
\footnote{http://www.atnf.csiro.au/research/pulsar/catalogue/}. Pulsar
distances are estimated by using the new electron density model of NE2001
\cite{cl02}. To make the consistency with $R_0=8$ kpc, all distances have
been scaled with a factor of 8/8.5=0.94. We binned these pulsars into four
groups according to their characteristic ages and obtained the scale-height
$h_g$ of each group through Gaussian fitting (filled circles in
Figure~\ref{hg_Norm_NewD}).  We then checked the relation between $h_g$ and
ages, assuming that the true age $t$ is equal to the characteristic age 
$\tau$. The resulting 1D velocity dispersion is $29\pm12$~km~s$^{-1}$, 
too small compared with previous results of several hundreds from proper 
motion measurement analyses (e. g. Lyne \& Lorimer 1994; Hansen \& Phinney
1997\nocite{ll94,hp97}). This indicates strong selection effects in the
assembled pulsar sample, as we justify below.

For young pulsars (t$<$8 Myr), $z=z_{\rm ini}+v_{\rm zini} t$.  If 
$z_{\rm ini}<\sim0.06$~kpc is ignored, a pulsar with $z$-velocity 
of 400~km~s$^{-1}$ will reach 0.4~kpc at 1~Myr, but 4~kpc at 10~Myr.
Previous pulsar surveys, like the Parkes multibeam survey which
discovered about half of all known pulsars, mainly covered $|b|\leq5^\circ$.
If the average pulsar distance is about 6~kpc, all pulsars of
$|b|\le5^\circ$  would have a $|z|<0.5$~kpc  at or nearer than
this average distance. Only pulsars with $\tau<8$~Myr and
a $z$-velocity less than  $v_{\rm zini}\approx z/t=0.5{\rm kpc}/8{\rm Myr}=$
63~km~s$^{-1}$ can be picked up in such a survey. The detected older
pulsars must have much smaller $z$-velocities, otherwise it have
to be much farther away than a distance of 6~kpc.

Pulsars younger than 1~Myr cannot move too far away even with a large
initial $z$-velocity because of their small ages. As a result,
the combined sample
would not suffer much selection effects due to survey regions. This is why
Lyne \& Lorimer (1994) chose 1 Myr as the cut-off age of the sample. Now we
try to use much more pulsars of $t<1$ Myr to estimate $z$-velocity. About
244 pulsars have been divided into three bins, with roughly the same
number of pulsars each. The 1D velocity dispersion in $z$ direction derived
from the scale-heights of $z$ distributions of three sub-samples is
175$\pm$56~km~s$^{-1}$, as shown in Figure~\ref{hg_Norm_NewD}. This
corresponds to a 3D velocity dispersion of $303\pm97$~km~s$^{-1}$ or a
mean velocity of $280\pm96$~km~s$^{-1}$.

To compare with previous estimates of velocities, we should use Taylor \&
Cordes~\shortcite{tc93} model for pulsar distance, and should not include
the pre-scaling factor from $R_0$. We then found the 1D velocity dispersion
of 187$\pm$64~km~s$^{-1}$, or a 3D velocity dispersion of
324$\pm$110~km~s$^{-1}$ and the mean velocity of $299\pm102$~km~s$^{-1}$.
These values are consistent with estimates given by Hansen \& Phinney
\shortcite{hp97}, the mean pulsar velocity of 250~km~s$^{-1}$, or by
Lyne \& Lorimer (1994), the mean of $450\pm90$~km~s$^{-1}$.

\subsection{Initial Velocity Dispersion of MSPs}

MSPs have a more complex evolutionary history than normal pulsars.
Generally, MSPs are old neutron stars spun up by mass and momentum
transferring from the companions (e.g.~Alpar et al.~\shortcite{acrs82}).
Their ``birth'' velocity is either the velocity of the binary system or its
sole velocity after the disruption of the binary system.  Another
consecutive problem is their ages. Characteristic ages from period and
period derivative probably do not reflect their true ages. Are they
chronologically old? Hansen \& Phinney~\shortcite{hp97} demonstrated 
that pulsars older than 10$^7$ yr show the asymmetric drifts. Toscano et 
al.~\shortcite{tsb+99} confirmed the asymmetric drifts of MSPs, which justify 
that MSPs are really dynamically old enough to be virialized. 
There is no doubt that MSPs are really old objects in the Milky Way. 

\begin{figure}
\psfig{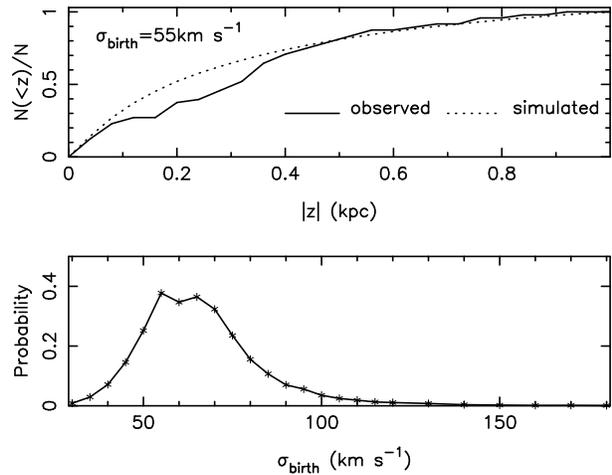}
\caption[]{K-S test: Cumulative probability distributions of simulated
and observed samples (upper panel), and the significance level for different 
initial velocities (lower panel).}
\label{hg_MSP_NewD}
\end{figure}

From our simulation, the height distribution of old pulsars
(e.g.~$t>$200~Myr) is stabilized. MSPs of all ages follow the same height
distribution. For the comparison of $z$-distributions of simulated sample
and observed sample, we did not take into account selection effects for the
MSP discovery. The observed sample consists of 48 MSPs from the
latest pulsar catalog after discarding the MSPs in globular clusters. The
simulated sample are old pulsars within 3~kpc from the Sun in a number of
sets of simulations with initial 1D $z$-velocity dispersion from
30~km~s$^{-1}$ to $180$~km~s$^{-1}$ (with a step of 5 km~s$^{-1}$).  The K-S
test is employed to check whether these two distributions are from the same
parent distribution. We find that $\sigma_{\rm birth}=60\pm10$~km~s$^{-1}$
gives the largest probability. If the velocities of MSPs follow a Maxwellian
distribution, the 1D velocity dispersion is most probably 60$\pm$10~km~s$^{-1}$
or the mean velocity dispersion most probably 96$\pm$16~km~s$^{-1}$, consistent
with previous results (e.g. Lyne et al. 1998).

\section{Conclusions}
We presented a generalized statistical picture of how pulsars move in our
Galaxy. The potential given by Paczy\'nski~(1990) can be a good
representation of mass distribution of our Galaxy, and has a simple
analytic formula. We found that the final height distributions are not 
sensitive to the forms of initial $z$ and $R$ distributions. The height 
distribution can be well fitted by a Gaussian function within $\sim$8~Myr, 
and the scale-height increases linearly with time. After that an extra 
exponential function is required to fit the height distribution.
The height distribution gets stabilized after about 200~Myr.

The height distribution of pulsars younger than 1~Myr implies directly that
the mean initial velocity of 280$\pm$96~km~s$^{-1}$. Comparison of the 
simulated sample of millisecond pulsars and observed sample suggests
the 1D initial velocity dispersion of MSPs to be most probably 
60$\pm$10~km~s$^{-1}$, consistent with estimates given by previous authors.

\section*{Acknowledgments}
We would like to thank the referee, Dr. Dunc Lorimer, for very detailed
comments and suggestions, which led to significant improvements of the paper.
We thank Prof. Andrew Lyne for valuable discussions and 
Drs. Walter Dehnen and Xianghua Li for their patient help on 
software. This work is supported by the National Natural Science 
Foundation of China (19903003 and 10025313) and the National Key Basic 
Research Science Foundation of China (G19990754) as well as the partner group 
of MPIfR at NAOC.


\begin{thebibliography}{}
\bibitem[\protect\citename{Alpar et al. }1982]{acrs82}
Alpar M. A., Cheng A. F., Ruderman M. A., Shaham J., 1982, Nat., 300, 728

\bibitem[\protect\citename{Arnaud \& Rothenflug }1981]{ar81}
Arnaud M., Rothenflug R., 1981, A\&A, 103, 263

\bibitem[\protect\citename{Arzoumanian et al. }2002]{acc02}
Arzoumanian Z., Chernoff D. F., Cordes J. M.,  2002, ApJ, 568, 289

\bibitem[\protect\citename{Bahcall }1986]{bah86}
Bahcall J. N., 1986, ARA\&A 24, 577

\bibitem[\protect\citename{Bailes \& Kniffen }1992]{bk92}
Bailes M., Kniffen D. A., 1992, ApJ, 391, 659

\bibitem[\protect\citename{Bhattacharya et al. }1992]{bwh+92}
Bhattacharya D., Wijers R. A. M. J., Hartman J. W., Verbunt F., 1992,
 A\&A, 254, 198

\bibitem[\protect\citename{Burton \& Gordon }1978]{bg78}
Burton W. B., Gordon M. A., 1978, A\&A, 63, 7

\bibitem[\protect\citename{Carlberg \& Innanen }1987]{ci87}
Carlberg R., Innanen K., 1987, AJ, 94, 666

\bibitem[\protect\citename{Cordes \& Chernoff }1997]{cc97}
Cordes J. M., Chernoff D. F., 1997, ApJ, 482, 971

\bibitem[\protect\citename{Cordes \& Chernoff }1998]{cc98}
Cordes J. M., Chernoff D. F., 1998, ApJ, 505, 315

\bibitem[\protect\citename{Cordes \& Lazio }2002]{cl02}
Cordes J. M., Lazio T. J. W., 2002, in press, astro-ph/0207156 

\bibitem[\protect\citename{Dehnen \& Binney }1998]{db98}
Dehnen W., Binney J., 1998, MNRAS, 294, 429

\bibitem[\protect\citename{Dewey \& Cordes }1987]{dc87}
Dewey R. J., Cordes J. M., 1987, ApJ, 321, 780

\bibitem[\protect\citename{Ferri\`{e}re }1998]{fer98}
Ferri\`{e}re K., 1998, ApJ, 497, 759

\bibitem[\protect\citename{Gonthier et al. }2002]{gob+02}
Gonthier P. L., Ouellette M. S., Berrier J. et al., 2002, ApJ, 565, 482

\bibitem[\protect\citename{Gunn \& Ostriker }1970]{go70}
Gunn J. E., Ostriker J. P., 1970, ApJ, 160, 979

\bibitem[\protect\citename{Hansen \& Phinney }1997]{hp97}
Hansen B.M.S., Phinney E.S., 1997, MNRAS, 291, 569

\bibitem[\protect\citename{Hartman et al. }1997]{hbw+97}
Hartman J. W., Bhattacharya D., Wijers R., Verbunt F., 1997, A\&A, 322, 477

\bibitem[\protect\citename{Hartman \& Verbunt }1995]{hv95}
Hartman J. W., Verbunt F., 1995, A\&A, 296, 110

\bibitem[\protect\citename{Helfand \& Tademaru }1977]{ht77}
Helfand D. J., Tademaru E., 1977, ApJ, 216, 842

\bibitem[\protect\citename{Itoh \& Hiraki }1994]{ih94}
Itoh N., Hiraki K., 1994, ApJ, 435, 784

\bibitem[\protect\citename{Johnston }1994]{joh94}
Johnston S., 1994, MNRAS, 268, 595

\bibitem[\protect\citename{Kuijken \& Gilmore }1989]{kg89}
Kuijken K., Gilmore G., 1989, MNRAS, 239, 571, 605

\bibitem[\protect\citename{Lorimer }1995]{lor95}
Lorimer D. R., 1995, MNRAS, 274, 300

\bibitem[\protect\citename{Lorimer }2004]{lor04}
Lorimer D. R., 2004, In: F. Camilo \& B. M. Gaensler, eds., IAU Symposium, 
Vol. 218, Young Neutron Stars and Their Environments.

\bibitem[\protect\citename{Lorimer et al. }1993]{lbd+93}
Lorimer D. R., Bailes M., Dewey R. J., Harrison P. A., 1993, MNRAS, 263, 403

\bibitem[\protect\citename{Lorimer et al. }1997]{lbh97}
Lorimer D. R., Bailes M., Harrison P. A., 1997, MNRAS, 289, 592

\bibitem[\protect\citename{Lyne \& Lorimer }1994]{ll94}
Lyne A. G., Lorimer D.R., 1994, Nat, 369, 127

\bibitem[\protect\citename{Lyne et al. }1998]{lml+98}
Lyne A. G., Manchester R. N., Lorimer D. R. et al., 1998, MNRAS, 295, 743

\bibitem[\protect\citename{Lyne, Manchester \& Taylor }1985]{lmt85}
Lyen A. G., Manchester R. N., Taylor J. H., 1985, MNRAS, 213, 613

\bibitem[\protect\citename{Ma\'iz-Apell\'aniz }2001]{mai01}
Ma\'iz-Apell\'aniz J., 2001, AJ, 121, 2737

\bibitem[\protect\citename{Manchester et al. }2004]{mhth04}
Manchester R. N., Hobbs, G., Teoh, A., Hobbs, M., 2004, 
In: F. Camilo \& B. M. Gaensler, eds., IAU Symposium, 
Vol. 218, Young Neutron Stars and Their Environments.

\bibitem[\protect\citename{Mihalas \& Binney }1981]{mb81}
Mihalas D., Binney J., 1981, Galactic Astronomy (San Francisco: Freeman)

\bibitem[\protect\citename{Miyamoto \& Nagai }1975]{mn75}
Miyamoto M., Nagai R., 1975, PASJ, 27, 533

\bibitem[\protect\citename{Mukherjee \& Kembhavi }1997]{mk97}
Mukherjee S., Kembhavi A., 1997, ApJ, 489, 928

\bibitem[\protect\citename{Narayan }1987]{nar87}
Narayan R., 1987, ApJ, 319, 162

\bibitem[\protect\citename{Narayan \& Ostriker }1990]{no90}
Narayan R., Ostriker J. P., 1990, ApJ, 352, 222

\bibitem[\protect\citename{Paczy\'{n}ski }1990]{pac90}
Paczy\'{n}ski B., 1990, ApJ, 348, 485

\bibitem[\protect\citename{Press et al. }1992]{ptv+92}
Press W. H., Teukolsky S. A., Vetterling W. T., Flannery B. P., 1992, 
Numerical Recipes in Fortran, Second edition. p.705

\bibitem[\protect\citename{Sofue \& Rubin }2001]{sr01}
Sofue Y., Rubin V., 2001, ARA\&A 39, 137

\bibitem[\protect\citename{Sofue et al. }1999]{sth+99}
Sofue Y., Tutui Y., Honma M. et al., 1999, ApJ, 523, 136

\bibitem[\protect\citename{Taylor \& Cordes }1993]{tc93}
Taylor J. H., Cordes J. M., 1993, ApJ, 411, 674

\bibitem[\protect\citename{Toscano et al. }1999]{tsb+99}
Toscano M., Sandhu J. S., Bailes M. et al., 1999, MNRAS, 307, 925

\bibitem[\protect\citename{Westerbout }1976]{wes76}
Westerbout G.: 1976, Maryland-Bonn Galactic 21-cm Line Survey, University of 
Maryland, College Park.

\end{thebibliography}
\end{document}